\documentclass[%
 aip,
 amsmath,amssymb,
 reprint,%
]{revtex4-1}

\usepackage{graphicx}% Include Fig. files
\usepackage{dcolumn}% Align table columns on decimal point
\usepackage{bm}% bold math
\usepackage[utf8]{inputenc}
\usepackage[T1]{fontenc}
\usepackage{mathptmx}
\usepackage{etoolbox}
\usepackage{siunitx}
\makeatletter
\def\@email#1#2{%
 \endgroup
 \patchcmd{\titleblock@produce}
  {\frontmatter@RRAPformat}
  {\frontmatter@RRAPformat{\produce@RRAP{*#1\href{mailto:#2}{#2}}}\frontmatter@RRAPformat}
  {}{}
}%
\makeatother
\begin{document}

\preprint{AIP/123-QED}

\title[Controlled epitaxy of room-temperature quantum emitters in gallium nitride]{Controlled epitaxy of room-temperature quantum emitters in gallium nitride}
% Force line breaks with \\

\author{Katie M. Eggleton}
\affiliation{School of Engineering, Cardiff University, Queen's Buildings, The Parade, Cardiff, CF24 3AA, United Kingdom}
  \affiliation{Translational Research Hub, Maindy Road, Cardiff, CF24 4HQ, United Kingdom}
\author{Joseph K. Cannon}
\affiliation{School of Engineering, Cardiff University, Queen's Buildings, The Parade, Cardiff, CF24 3AA, United Kingdom}
  \affiliation{Translational Research Hub, Maindy Road, Cardiff, CF24 4HQ, United Kingdom}
\author{Sam G. Bishop}
\affiliation{School of Engineering, Cardiff University, Queen's Buildings, The Parade, Cardiff, CF24 3AA, United Kingdom}
  \affiliation{Translational Research Hub, Maindy Road, Cardiff, CF24 4HQ, United Kingdom}
   \author{John P. Hadden}
\affiliation{School of Engineering, Cardiff University, Queen's Buildings, The Parade, Cardiff, CF24 3AA, United Kingdom}
  \affiliation{Translational Research Hub, Maindy Road, Cardiff, CF24 4HQ, United Kingdom}
\author{Chunyu Zhao}
 \affiliation{Department of Materials Science and Metallurgy, University of Cambridge
, 27 Charles Babbage Rd, Cambridge CB3 0FS, United Kingdom}
 \author{Menno J. Kappers}
 \affiliation{Department of Materials Science and Metallurgy, University of Cambridge
, 27 Charles Babbage Rd, Cambridge CB3 0FS, United Kingdom}
 \author{Rachel A. Oliver}
 \affiliation{Department of Materials Science and Metallurgy, University of Cambridge
, 27 Charles Babbage Rd, Cambridge CB3 0FS, United Kingdom}
\author{Anthony J. Bennett} 
 \affiliation{School of Engineering, Cardiff University, Queen's Buildings, The Parade, Cardiff, CF24 3AA, United Kingdom}
  \affiliation{Translational Research Hub, Maindy Road, Cardiff, CF24 4HQ, United Kingdom}
 \email{BennettA19@cardiff.ac.uk}

\date{\today}% It is always \today, today,

\begin{abstract}
The ability to generate quantum light at room temperature on a mature semiconductor platform opens up new possibilities for quantum technologies. Heteroepitaxial growth of gallium nitride on silicon substrates offers the opportunity to leverage existing expertise and wafer-scale manufacturing, to integrate bright quantum emitters in this material inside cavities, diodes and photonic circuits. Until now it has only been possible to grow GaN quantum emitters at uncontrolled depths on sapphire substrates, which is disadvantageous for potential device architectures. Here we report a method to produce GaN quantum emitters by metal-organic vapor phase epitaxy at a controlled depth in the crystal through application of a silane treatment and subsequent growth of 3D islands. We demonstrate this process on highly technologically relevant silicon substrates, producing room-temperature quantum emitters with a high Debye-Waller factor and strongly anti-bunched emission.

\end{abstract}

\maketitle

\section{\label{sec:intro}Introduction}

Solid-state quantum emitters (QEs) are essential components for emerging quantum technologies \cite{Atature2018, awschalomQuantumTechnologiesOptically2018}, such as quantum communications \cite{bhaskarExperimentalDemonstrationMemoryenhanced2020}, quantum sensors \cite{Aslam2023} and quantum computing \cite{kimbleQuantumInternet2008, childressDiamondNVCenters2013}. There is a growing interest in room-temperature QEs that can be implemented in practical and scalable materials. Quantum emitters based on impurity-vacancy complexes in diamond \cite{Kurtsiefer2000,Childress2013, Hepp2014}, hexagonal boron nitride \cite{Gottscholl2020, Plo2025}, silicon carbide \cite{Klimov2014,Castelletto2014} and silicon \cite{johnstonCavitycoupledTelecomAtomic2024,Durand2021, beaufilsOpticalPropertiesEnsemble2018,Higginbottom2022} have been extensively studied, and many have demonstrated anti-bunching at room temperature. 

Wurtzite nitrides are a promising material platform for quantum technology owing to the extensive existing knowledge of their epitaxy and a proven array of device architectures developed for commercial applications in lighting and power electronics. QEs in gallium nitride (GaN) \cite{Berhane2017,Bishop2022} and aluminum nitride (AlN) \cite{Xue2020, Bishop2020} are optically stable and emit across a wide spectral range (from visible to near-IR). GaN has demonstrated bright QEs with a high fraction of emission occurring through the zero-phonon line \cite{Bishop2022} as quantified by the Debye-Waller (DW) factor. Some GaN QEs have been reported to emit at telecommunications wavelengths \cite{Zhou2018} though it is not known if these have the same origin as QEs seen in the \SI{600}{} to \SI{900}{\nano \meter} range. Recent reports have shown that QEs in GaN have spin-linked energy levels that make them amenable to optically detected magnetic resonance spectroscopy \cite{Luo2024, Eng2025}. Their high brightness and single-photon purity has also led to GaN QEs being integrated into prototype quantum communication links \cite{Zhang2025}.

\begin{figure*}[t]
    \centering
    \includegraphics[scale=1]{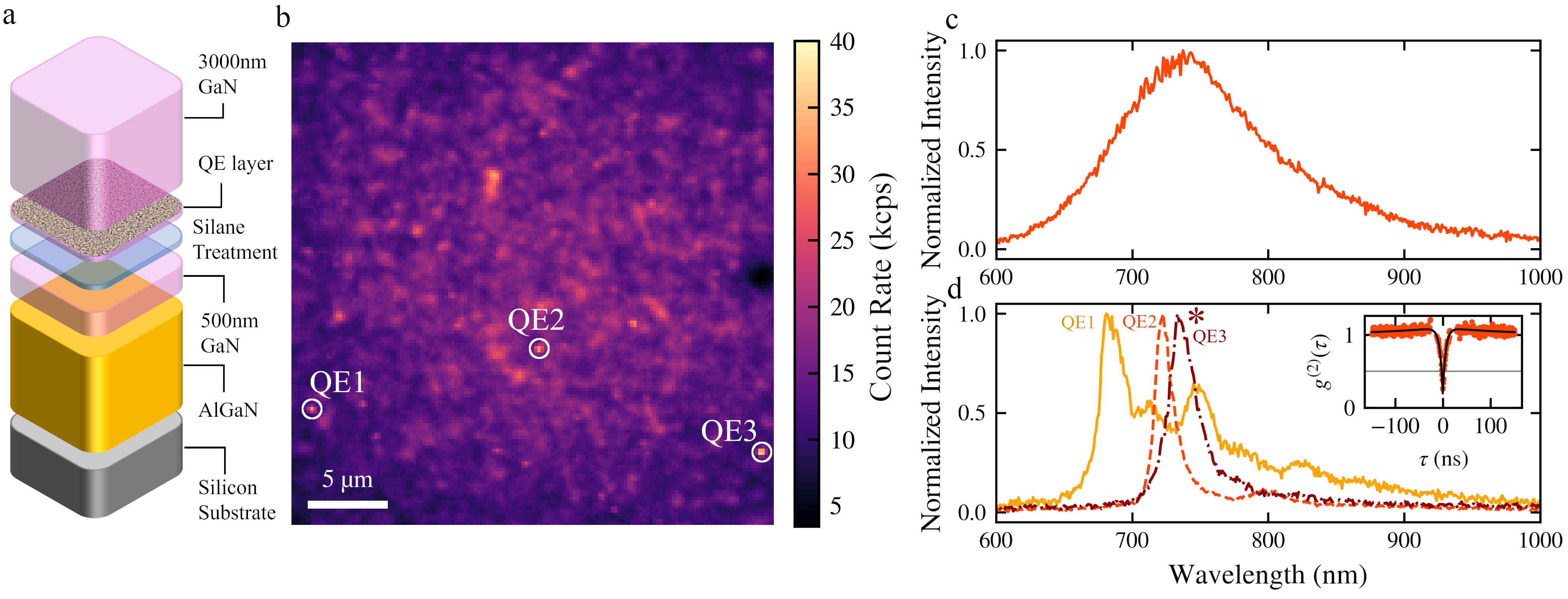}
    \caption{Controlled growth of GaN quantum emitters on silicon. a) Illustration of the sample stack. b) Confocal fluorescence scan map of sample, collected in the spectral range 550 to \SI{1000}{nm}. c, d) Room-temperature photoluminescence spectra of a typical AlN QE in this sample and the GaN QEs, marked in b), respectively. The inset in d) is a second-order correlation measurement of QE3, marked by an asterisk in d).}
    \label{fig1}
\end{figure*}

The dominant commercial technology for growth of III-nitrides is metal-organic vapor phase epitaxy (MOVPE) on sapphire, silicon carbide or silicon substrates \cite{Krost2002, Zhu2013}, the latter of which are commercially available on \SI{300}{\milli\meter} wafers. Owing to the `melt-back' etching of silicon (111) by the gallium precursor \cite{Zhu2013}, the epilayer is often initiated with an AlN layer before gradually increasing the Ga-content of the film. 

We have previously shown that AlN growth on Si results in emission from single AlN QEs with phonon-broadened spectra extending from \SI{600}{} to \SI{900}{\nano\meter}. At room temperature, these AlN QEs have a low Debye-Waller factor \cite{Cannon2024AlNSilicon} and unknown spin-physics. We have also shown that QEs in AlN epilayers, grown on both sapphire and silicon substrates, preferentially form in the first few tens of nm of AlN growth \cite{yagciTrackingCreationSingle2024a}. This corresponds to the formation of QEs in a highly dislocated and unintentionally doped initial layer. It would be  advantageous to grow GaN QEs in a controlled manner and to achieve precise control over the depth of formation, enabling integration in optical cavities, opto-electronic devices and quantum sensing at surfaces.

In this work, we report an epitaxial process for the low-density growth of QEs at a controlled depth away from the highly-defective epilayer-substrate interface region by mimicking the conditions in which GaN QEs form at the sapphire-GaN interface. We demonstrate this process via growth of stable, high DW-factor GaN QEs emitting at room temperature on silicon. Correlative optical intensity and micro-structural data reveals QEs located in the sparse 3D islands of GaN grown on a treated GaN surface. We also show that overgrowth of these QEs with a planar layer of GaN, which is essential for device applications, preserves the buried QEs. We confirm the GaN QEs are located only at the intended depth within the GaN layer via controlled etching of the epitaxial stack. Thus, we report that integration of QEs in GaN-on-silicon is possible with precise control of their depth, offering an attractive platform for room-temperature quantum light sources and sensors.

\section{Results}

Figure \ref{fig1} shows a sample design we have developed to grow QEs at arbitrary depth in a GaN layer. We employ a \SI{1}{\milli\meter} thick n-type Si (111) substrate of \SI{150}{\milli\meter} diameter on which epitaxy is initialized with the growth of a nominal thickness of approximately \SI{250}{\nano\meter} AlN. Growth proceeds with a compositionally-graded \SI{1.7}{\micro\meter} AlGaN layer with increasing Ga content, then a \SI{1.0}{\micro\meter} GaN buffer layer (total epitaxy height of \SI{3.0}{\micro\meter}). This GaN-on-Si pseudo-substrate was then cleaved into 3$\times$\SI{3}{\centi\meter\squared} pieces and the regrowth on each piece was started with a \SI{500}{\nano\meter} GaN connecting layer. At this point the GaN surface was treated with silane (SiH\textsubscript{4}) and ammonia (NH\textsubscript{3}) for \SI{720}{\second} which forms a thin layer of SiN\textsubscript{x} similar to what is created during the initial nitridation of sapphire before GaN growth \cite{Yadda2018, Haffouz1999}. A GaN layer equivalent to 10 nm thickness, which form islands in the SiNx layer pinholes, is then deposited at low temperature and annealed in ammonia at 1020 C. We show here this layer contains QEs. Finally, this layer is overgrown with \SI{3.0}{\micro\meter} of GaN to prevent QE instability from reactions with air and surface charges. \SI{3.0}{\micro\meter} was used to ensure a planarized surface. There are approximately $10^{-3}$ emitters per \SI{10}{\nano\meter} of spectral bandwidth per square micron. The completed epilayer stack is shown schematically in Fig. \ref{fig1}(a). 

Room-temperature confocal laser-scanning photoluminescence (PL) measurements of the complete epitaxial stack reveal a distribution of point-like emitters, as highlighted in Fig. \ref{fig1}(b). Emitters with an optical fingerprint consistent with AlN-on-silicon QEs \cite{Cannon2024AlNSilicon}, appear at a density of $\approx \SI{1}{\micro \meter^{-2}}$. An example spectrum is shown in Fig. \ref{fig1}(c). We have previously confirmed by studying a sequence of samples that terminate after the various stages of the growth sequence that these QEs are formed in the initial stages of the AlN layer. QEs with narrower spectra, consistent with GaN QEs \cite{Berhane2017,Bishop2022}, appear at an almost 2 orders-of-magnitude lower density. As with previous reports, each emitter has a well-defined emission dipole \cite{Berhane2017,Bishop2022}. We confirm later in this report that the GaN QEs are formed in the nominally \SI{10}{\nano\meter} layer that immediately follows the silane treatment.  

\begin{figure*}[t]
    \centering
    \includegraphics[scale = 1]{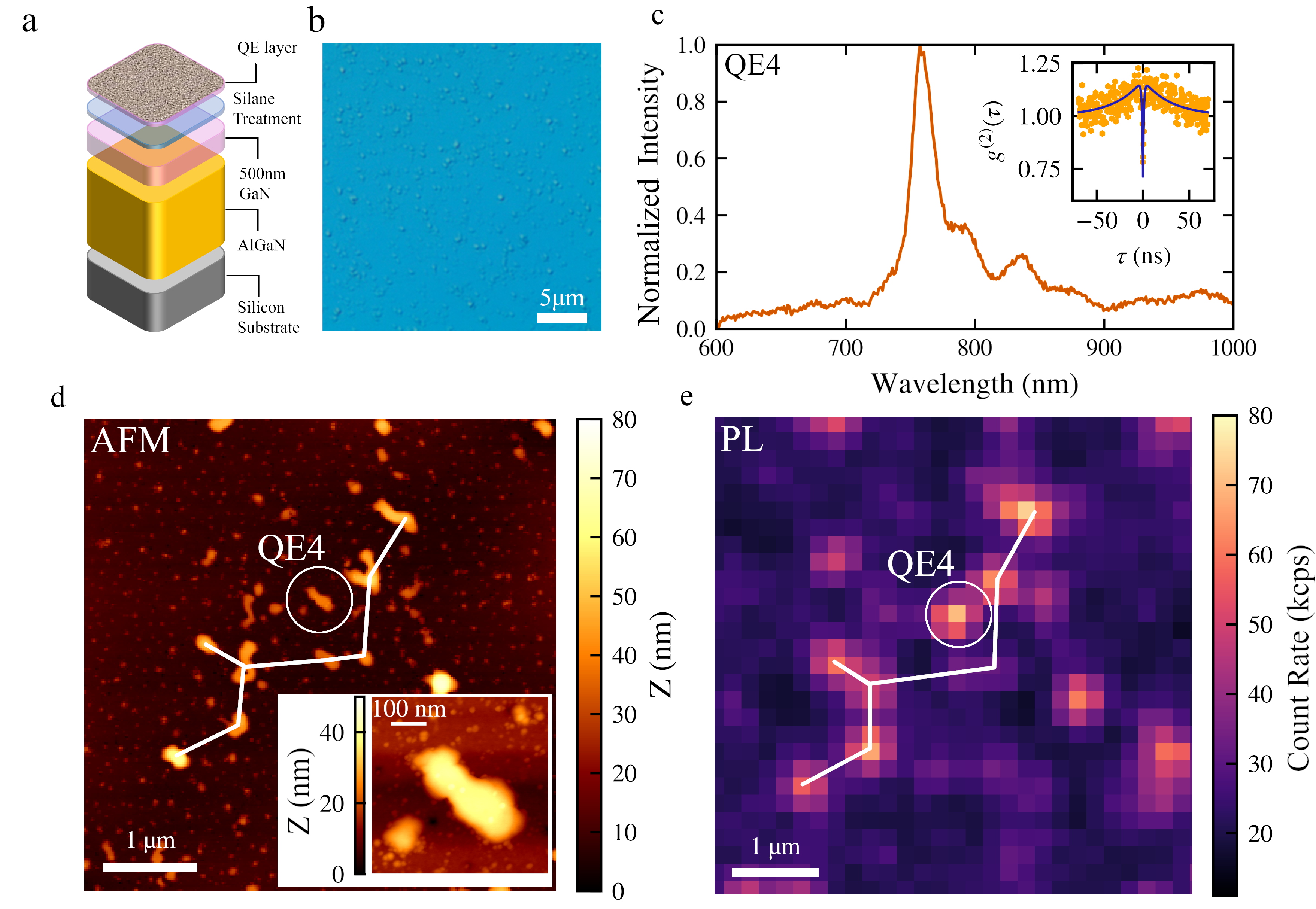}
    \caption{Emission from uncapped GaN nano-islands grown directly on the silane-treated surface. a) Illustration of the sample stack. b) Optical micrograph. c) Photoluminescence spectrum of QE4, which is highlighted in d, e). The second-order correlation of QE4 is shown in the inset. d) Atomic force micrograph image of the sample surface and e) corresponding confocal scan map. The inset in d is a smaller scale image of QE4. The white line linking some islands in d and e is provided as a guide to the eye.}
    \label{fig2}
\end{figure*}

The PL spectra of three typical GaN QEs are presented in Fig. \ref{fig1}(d). Each spectrum shows a prominent zero-phonon line (ZPL), with linewidths ranging from \SI{19}{} to \SI{33}{\nano\meter}, and DW-factors of \SI{0.31}{} to \SI{0.63}{}, consistent with other studies of GaN QEs \cite{Bishop2022, Berhane2017a}. The GaN QEs all display antibunching in their room temperature emission under CW laser excitation, indicative of quantized energy states. A second-order correlation, measured across the full spectrum of QE3 using a Hanbury-Brown and Twiss interferometer at \SI{90}{\micro \watt} excitation power, is shown in the inset of Fig. \ref{fig1}(d). The correlation shows both bunching and antibunching at room temperature and was fit using an empirical model \cite{Cheng2025, Patel2022, guoEmissionDynamicsOptically2024}, 

\begin{equation}
    g^{(2)}(\tau) = 1 + \alpha e^{-|\tau|/\tau_1} + \beta e^{-|\tau|/\tau_2},
    \label{eq:HBT}
\end{equation}

where $\tau_{1,2}$ are the antibunching and bunching timescales and $\alpha, \beta$ are the antibunching and bunching amplitudes, respectively. The antibunching and bunching timescales are related to the rates associated with the ground, excited and shelving state of the model. For QE3, $g^{(2)}(\tau =0) = 0.26 \pm 0.05$ which is below the conventional limit of $g^{(2)}(0)=0.5$ at zero delay for a single-emitter. The antibunching and bunching timescales are \SI{5.6\pm0.2}{} and \SI{1980\pm60}{\nano\second}, respectively. The antibunching and bunching amplitudes are $\SI{-0.92\pm 0.02}{}$ and $\SI{+0.11\pm 0.01}{}$ respectively. The value of $g^{(2)}(0)$ is limited by the broad-band background emission from the AlN QEs near the AlN-Si interface.

To understand the growth of the GaN QEs on the silane-treated surface we now present data on a second sample. The sample stack is illustrated in Figure \ref{fig2}(a). The sample was removed from the growth reactor after the anneal of the GaN islands, whose formation was induced by the silane treatment, labeled ``QE layer". A Nomarski microscope image of the sample reveals a rough, islanded surface (Fig. \ref{fig2}(b)). Surprisingly, this sample displays GaN QE emission even though the formed emitters are expected to be within tens of nanometers of the surface. A typical spectrum and second-order correlation measurement from one of these QEs is shown in Fig. \ref{fig2}(c). The proximity of the emitter to the surface impedes our ability to fully suppress fluorescent background, resulting in $g^{(2)}(0) = \SI{0.70 \pm 0.07}{}$, consistent with the presence
of a quantum emitter and a significant background emission. 

\begin{figure*}[t]
    \centering
    \includegraphics[scale = 1]{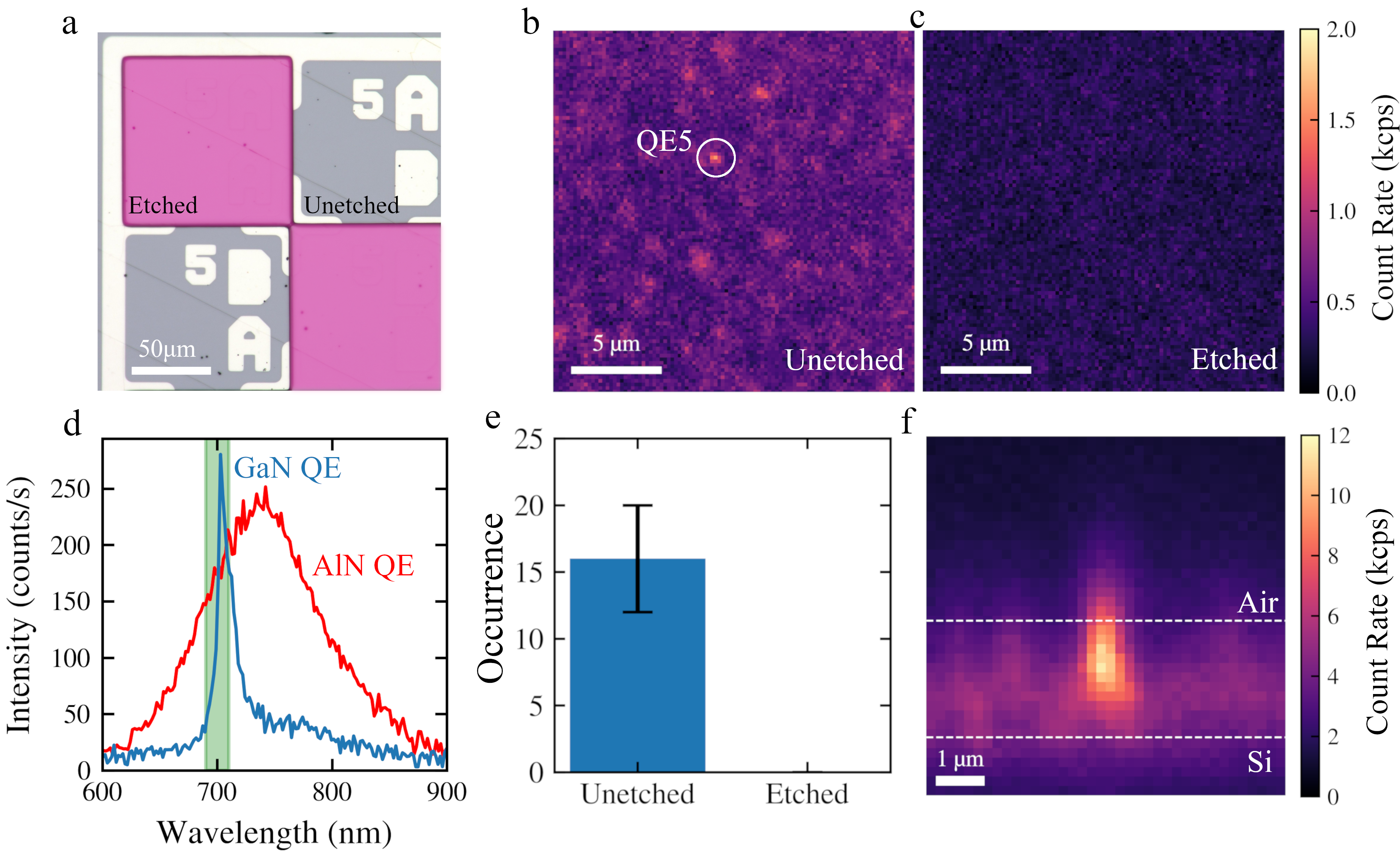}
    \caption{Patterned sample etching and mapping. a) Optical image of the etched sample surface showing a checkerboard pattern of labeled etched squares. b, c) Confocal scan maps of unetched/etched regions, respectively. d) Typical room-temperature GaN and AlN QE spectra. e) Bar chart showing the number of GaN QEs identified across all test areas. f) Depth resolved confocal scan map of QE5. The air-GaN and AlN-Si interfaces are marked with a dotted line.}
    \label{fig3}
\end{figure*}

We compare the topology of the GaN islands from atomic force microscopy (AFM, Fig. \ref{fig2}(d)) with features in the PL scan map (Fig. \ref{fig2}(e)). We observe a clear correlation between the islands of GaN material and points of increased PL intensity. Although we use spectral filters to eliminate laser bleed-through, Raman scattering and yellow-band luminescence, and the confocal arrangement to minimize emission from AlN QEs at the Si-interface, we cannot rule out the possibility that this spatial pattern is caused by a lensing effect from the nano-islands, locally enhancing some weak background photoluminesence emission. For the majority of the island features in the AFM image, the corresponding PL emission does not display any structure in its spectrum or demonstrate anti-bunching, with highlighted QE4 being an exception.  

The AFM images demonstrate a distribution of island sizes ranging from tens of nanometers in width up to several hundreds of nanometers. Emission from the smaller islands which are visible at high density in the AFM image, cannot be resolved in the PL map. Nevertheless, the AFM-PL correlative microscopy enables localization the emitter QE4 in Fig.  \ref{fig2}(c) within one of the larger elongated islands. We note that topologically similar GaN islands do not all contain QEs. 

To develop useful devices it will be necessary to embed the GaN QEs within flat epilayers at a fixed depth. This will allow creation of contacts, mirrors and doped layers at controlled distances from the QEs, and mitigate interactions with surface states. However, we have observed that the overgrowth on the silane-treated surface is not trivial and requires at least \SI{3}{\micro\meter} of GaN growth to fully coalesce and create a suitably flat c-plane surface. 

A \SI{3.2}{\micro \meter} etch was carried out to pattern a checkerboard of etched and unetched regions on the sample introduced in Fig. \ref{fig1}(a), photographed in Fig. \ref{fig3}(a). Etched areas are highlighted in pink, unetched areas contain unique labels in a surface metal layer. PL scanning measurements were performed across a total area of $\SI{12100}{\micro \meter \squared}$ on both etched and unetched regions. A $\SI{700\pm5}{\nano\meter}$ band-pass filter was used to suppress fluorescence of the AlN QEs and aid identification of GaN QEs within the pass band. Typical scan maps are shown in Fig. \ref{fig3}(b) and (c) for the unetched and etched regions, respectively with QE5 marking the location of a GaN QE. After etching, only AlN QEs remain which are heavily suppressed by the band-pass filtering, as illustrated in Fig. \ref{fig3}(d).  

In total 16 QEs were observed across the unetched regions within the \SI{10}{\nano\meter} spectral range, shown in Fig. \ref{fig3}(e). This corresponds to a GaN QE density \SI{130 \pm 30}{\milli\meter  \tothe{-2}} per \SI{}{\nano\meter} of spectrum. GaN QEs were not observed across the $\SI{12100}{\micro \meter \squared}$ etched regions, confirming they are located in the GaN layer lying above the silane surface treatment. 

% Fig. \ref{fig3}(f) shows the depth profile of a GaN QE highlighted in Fig. \ref{fig3}(b). The raw data is plotted in real-space units of sample position (multiple material indices refract the laser focal depth during the scan). Simultaneous measurement of the reflection from the sample allows determination of the air/GaN and AlN/Si interfaces, shown as dotted white lines. QE5 is located in the center of the epilayer, as expected. Background emission closer to the semiconductor-substrate interface is consistent with the presence of AlN QEs.

%Fig. \ref{fig3}(f) illustrates the depth profile of QE5 highlighted in Fig. \ref{fig3}(b). The measurement is taken scanning the laser position in X and the focus in Z, simultaneously measuring the fluorescence and the reflection of the laser from the sample. We do not correct the measurement position to account for refraction at the interfaces of the sample stack. Considering the interfaces visible in the reflection measurement, we determine that the emitter is located at the centre of the epilayer as expected.

Fig. \ref{fig3}(f) illustrates the depth profile of QE5 highlighted in Fig. \ref{fig3}(b). The measurement is taken scanning the laser position in X and the focus in Z, simultaneously measuring the fluorescence and the reflection of the laser from the sample. Spectral filtering was used to separate the laser reflection from the fluorescence signal. The labeled interfaces are determined from the corresponding reflections. We do not correct the position of the dashed lines to account for refraction. The measurement shows the emitter is located at the centre of the epilayer as expected.

\section{Discussion}

\begin{figure*}[t]
    \centering
    \includegraphics[scale = 1]{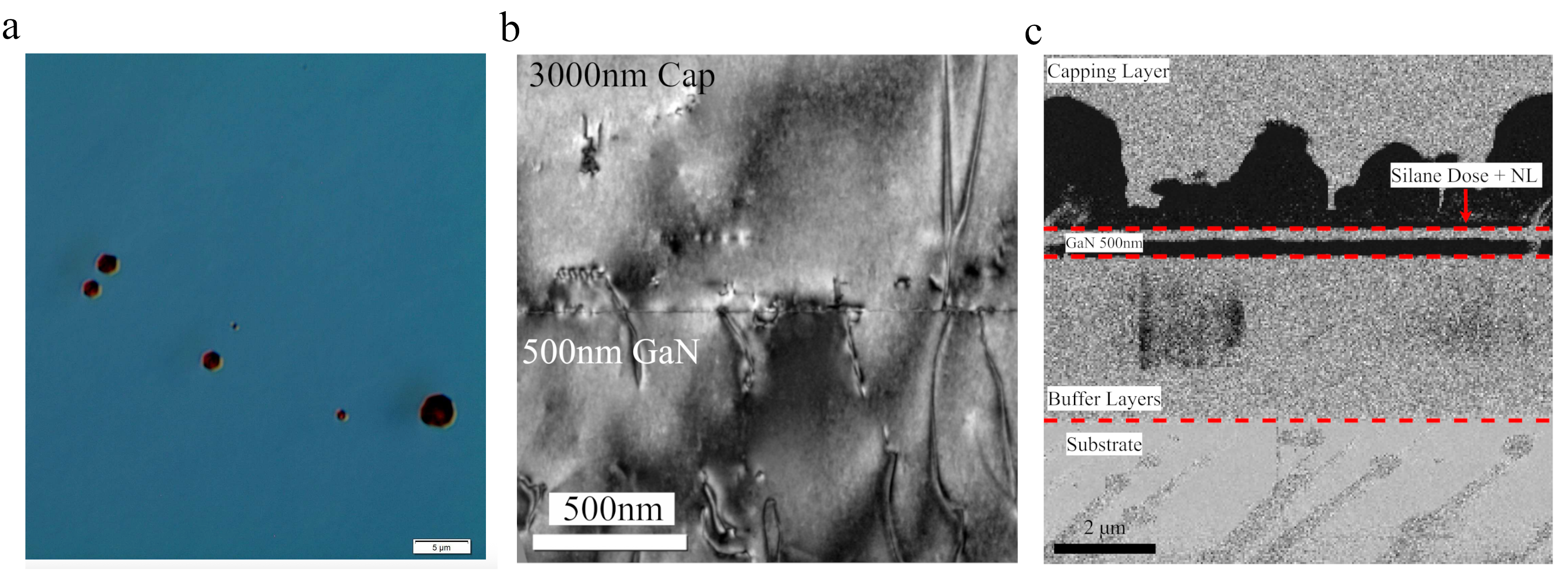}
    \caption{Microstructure of the capped sample. a) Optical micrograph of the sample after growth of a \SI{3.0}{\micro\meter} cap showing the presence of large voids. b) Cross-sectional transmission electron microscope image of the area around the silane-treated layer showing the presence of threading dislocations. c) Scanning capacitance microscopy phase-image of the epilayer. Darker regions illustrate the presence of n-type conductivity in the layer, most prominently in the continuous layer marking the start of \SI{500}{\nano\meter} GaN regrowth on the GaN-on-Si pseudo-substrate and the dark uneven layer directly above the silane treatment. }
    \label{fig4}
\end{figure*}

Having proven that GaN-QEs can be grown in controlled thin layers by MOVPE, it is important to assess if this result provides information on the origin of GaN-QEs. In the literature a number of proposals have been put forward \cite{Berhane2017, Nguyen2019, Li2020} but clear determination of their structure has yet to be made. It is known that GaN QEs are correlated with growth mode and preferentially form during growth on sapphire substrates \cite{Nguyen2019}. This first report of GaN QE growth on the silane treated GaN surface implies that these emitters can be grown inside GaN epilayers on any substrate. Structural characterization of the overgrown sample is presented in Figure \ref{fig4}. Fig. \ref{fig4}(a) shows a Nomarski microscope image of the surface revealing that, even after \SI{3.0}{\micro\meter} of GaN overgrowth, the film is not fully coalesced, with holes remaining in the surface. 

A bright field transmission electron microscope (TEM) image, taken on a vertical cross-section of the sample and shown in Fig. \ref{fig4}(b), reveals threading dislocations terminating in the region of the silane treatment. At this location a layer of $\mathrm{SiN_x}$ is present which initiates 3D island growth of GaN. Some of the threading dislocations bend and meet resulting in annihilation \cite{Kappers2010} in a similar manner to the growth of GaN on sapphire by the `3D-2D' approach \cite{Oliver2008}. TEM is sensitive to changes in lattice parameter and structural defects, but does not reveal the presence of dopants within the semiconductor. We therefore perform scanning capacitive microscopy on a cleaved cross section of the epilayer, and show the phase images in Fig. \ref{fig4}(c). The irregular dark regions directly above the $\mathrm{SiN_x}$ layer correspond to n-type dopant incorporation regions in the capping layer \cite{Oliver2009}. Below that region is a continuous dark stripe that marks the GaN regrowth interface in which unintentional doping is present \cite{Sumner2009}, which is a known phenomomenon of Si incorporation in III-nitride regrowth on an Aixtron MOVPE reactor. The lighter, noisy regions in the upper section of the GaN cap and AlGaN buffer correspond to insulating material where charge cannot deplete \cite{Oliver2009}. Unintentional n-doping is also observed in GaN on sapphire epitaxy \cite{Sumner2009}, where GaN QEs also form.

The spectral properties of the QEs highlighted in this work is summarised in Table \ref{tab:QE_Spectra}. We measure the QEs’ spectra with a detection window between \SI{600}{}-\SI{1000}{\nano\meter}. QEs at multiple different wavelengths, in addition to the ones presented in the manuscript were observed with ZPLs between \SI{640}{\nano\meter} to \SI{770}{\nano\meter} under \SI{520}{\nano\meter} laser excitation.

\begin{table}[h]
    \centering
    \begin{tabular}{|c|
        >{\centering\arraybackslash}p{1.2cm}|
        >{\centering\arraybackslash}p{1.5cm}|
        >{\centering\arraybackslash}p{1.8cm}|
        >{\centering\arraybackslash}p{1.9cm}|}
    \hline
    QE Label & Capping layer? & ZPL Wavelength (nm) & ZPL FWHM (nm) & Debye-Waller Factor \\
    \hline
    QE1 & Yes & 688.2 & 33.5 & 0.31 \\
    \hline
    QE2 & Yes & 724.0 & 19.5 & 0.58 \\
    \hline
    QE3 & Yes & 737.4 & 29.0 & 0.63 \\
    \hline
    QE4 & No  & 761.2 & 31.8 & 0.39 \\
    \hline
    QE5 & Yes & 695.7 & 27.6 & 0.30 \\
    \hline
    \end{tabular}
    \caption{Marked QE spectral properties}
    \label{tab:QE_Spectra}
\end{table}

Although these QEs appear in proximity to n-doping, there is no direct evidence that the QEs arise from a Si related vacancy specifically; however the possibility is not ruled out in this work. An alternate hypothesis is that, modifications to the Fermi energy locally provides an environment where these QE can become optically active.  Determining the microscopic origin of a given QE is challenging, but simulations have suggested several possible defect configurations in GaN \cite{Li2020, gaoIntrinsicDefectProperties2004, lyonsComputationallyPredictedEnergies2017, Berhane2017}. Further experimental work is required to confirm these attributions.

\section{Conclusion}
We demonstrate a process that enables controlled epitaxy of QEs at a defined depth in GaN, amenable to growth on sapphire or silicon substrates. These emitters have high DW factors and display strong antibunching at room temperature, even in close proximity to the surface of the epilayer. This controlled depth opens up a range of possibilities to control the QEs in heterostuctures that leverage the advanced technologies available in the III-nitrides system. Examples include the development of cavities with emitters located at the anti-node of the electric field. However, such devices will require the the planarization of the GaN capping layer, which data in Figure \ref{fig4} suggests will be non-trivial. Given sufficient control over the material doping, the development of diodes that allow electronic tuning, charging or injection of carriers in the QEs can be explored. The inclusion of microwave striplines that can allow for optically-detected magnetic resonance sensing could also be fabricated. Compatibility with mature manufacturing processes and growth on silicon makes these QEs viable candidates for future applications across quantum technologies.

% Experimental section

\section{Experimental Section}
\subsection{Optical Characterization}
Photoluminescence measurements were performed using a custom scanning confocal 4f microscope. A \SI{520}{\nano\meter} wavelength laser was used for all excitation. QE emission was isolated using a \SI{532}{\nano\meter} dichroic mirror and a \SI{550}{\nano\meter} long pass filter in the collection path. A linear polarizer was aligned to the emitter dipole to maximize the signal relative to the background. A 0.9 numerical aperture refractive objective was used to focus the laser onto the sample and collect fluorescence. Photon detection was performed using Excelitas fiber coupled SPCM-AQRH silicon avalanche photodiodes (APDs). The second order correlation in the inset of Fig. \ref{fig1}(d) is masked between delay times of $\pm \left( 15\rightarrow \SI{27}{\nano\second} \right)$ to omit counts from in-fiber reflections of detection flashes from the APDs. Spectral measurements using a \SI{328}{\milli\meter} spectrometer with an 80 lines per mm grating and a silicon CCD. The confocal microscope used for these experiments has a Rayleigh range of $\approx$ \SI{2.4}{\micro\meter}. \\

\subsection{Growth} 
All samples were grown by MOVPE. The epilayer was grown in an Aixtron 1$\times$6 inch close-coupled showerhead reactor with trimethylaluminum (TMAl), trimethylgallium (TMGa) and ammonia ($\mathrm{NH_3}$) as the precursors and hydrogen ($\mathrm{H_2}$) as the carrier gas. Emissivity-corrected pyrometry was used to control the substrate temperature. Before growth, the Si (111) substrate was heated inside the reactor at \SI{1050}{\celsius} to remove the native surface oxide in situ without any chemical treatment. This was followed with a change of temperature to \SI{950}{\celsius} for \SI{60}{\second} of $\mathrm{NH_3}$ predose followed by \SI{30}{\nano\meter} of AlN.  The temperature was then ramped to \SI{1080}{\celsius} for the growth of the remaining \SI{230}{\nano\meter} of AlN.  Next, the temperature was reduced to \SI{1030}{\celsius} for growth of a \SI{1.7}{\micro\meter} thick graded AlGaN layer with the group-III composition changing from 76\% Al to 25\%. A \SI{1.0}{\micro\meter} thick GaN buffer grown at \SI{1020}{\celsius} finished the pseudo-substrate growth process.  The regrowth of 3$\times$\SI{3}{\centi\meter} pseudo-substrate pieces was started with a slow temperature ramp to \SI{1020}{\celsius} in ammonia and hydrogen, before initiating a \SI{500}{\nano\meter} thick GaN `connecting' layer.  After cooling to \SI{860}{\celsius}, the GaN surface was treated with silane ($\mathrm{SiH_4}$) for \SI{720}{\second} in the presence of ammonia and hydrogen.  After cooling to \SI{540}{\celsius}, a \SI{10}{\nano\meter} thick GaN layer is grown and annealed in ammonia and hydrogen at \SI{1020}{\celsius}. Finally, this layer is overgrown with \SI{3.0}{\micro\meter} of GaN at the same temperature.
The process for QE growth is reproducible and has been implemented across a further 7 samples, with all of them containing QEs with similar properties to those presented in this work. As part of this study we have also investigated shorter silane dose times and found that times above 720s are required to observe GaN QEs. We also investigated a sample series using nominally 10, 20 and 40 nm GaN layers on silane-treated surfaces. Thicker layers resulted in coalesced islands which showed no quantum emission.   \\

\subsection{Microstructure Imaging}
Transmission electron microscopy in Fig. \ref{fig4}(b) was carried out by Integrity Scientific Ltd. Scanning capacitance microscopy in Fig. \ref{fig4}(c) was performed on a Bruker Dimension Icon Pro equipped with a Bruker SCM module. A Bruker SCM-PIC tip was employed.

\begin{acknowledgments}
We acknowledge financial support provided by EPSRC via Grant No. EP/X015300/1, EP/P024947/1, EP/R00661X/1, EP/T017813/1, and EP/X03982X/1. Device processing was carried out in the cleanroom of the ERDF-funded Institute for Compound Semiconductors (ICS) at Cardiff University.
\end{acknowledgments}

\section*{Data Availability Statement}

The data that support the findings of this study are openly available in Cardiff University Research Portal at http://doi.org/10.17035/cardiff.29617235.

\section*{Author Declarations}
\subsection*{Competing interests}
All authors declare they have no competing interests.

\section*{References}
\bibliography{references}% Produces the bibliography via BibTeX.

\end{document}